\documentstyle[mprocl,epsf]{article}

\newcommand{\dfrac}[2]{{\displaystyle\frac{#1}{#2}}}
\newcommand{\cfrac}[2]{\dfrac{\mathstrut #1}{#2}}
\def\DI{\Delta I}

\begin{document}

\title{Weak Decay of $\Lambda$ in Nuclei:\\ 
Direct Quark Mechanism vs Meson Exchange}

\author{Makoto Oka, Kenji Sasaki}

\address{Department of Physics, Tokyo Institute of Technology\\
        Meguro, Tokyo 152-8551 Japan\\E-mail: 
        oka@th.phys.titech.ac.jp}   
\author{Takashi Inoue\footnote{Present address: 
Department of Physics, Tokyo Institute of Technology,
Meguro, Tokyo 152-8551 Japan.}}
\address{Department of Physics, Univ.\ of Tokyo\\
        Hongo, Tokyo 112-0033 Japan}

\maketitle\abstracts{
Nonmesonic decays of $\Lambda$ in nuclear medium and 
light hypernuclei are studied
by using the $\Lambda N \to NN$ weak transition potential derived from
the meson exchange mechanism and the direct quark mechanism.
The long range part of the transition potential is described by
exchanges of the pseudoscalar mesons 
($\pi$, $K$, $\eta$), while the vector mesons ($\rho$, $\omega$, $K^\ast$)
may be considered as the medium- and short-range part in the meson 
exchange picture.
We propose the direct quark transition potential as the short range part, 
which is derived from the matrix elements of the $\Delta S=1$ 
effective weak Hamiltonian in the two baryon states.
The results indicate that the direct quark contribution is 
significantly large and its behavior is qualitatively different 
from the vector meson exchanges. 
We also find that the decay rate is sensitive to the choice of form factor
and that a soft cutoff must be used for the pion-baryon verteces
so that the strong tensor transition is suppressed.
We find that the $\pi + K + DQ$ results are compatible 
with experiment although the $n/p$ ratio is still too large.
The $\pi^+$ decays of light hypernuclei are related to the $\DI=3/2$ 
amplitudes of the nonmesonic decay.  The role of chiral symmetry for 
the pionic decays are discussed.}

\section{Introduction}

Mesonic decays of $\Lambda$ are suppressed in heavy hypernuclei
by the Pauli exclusion principle of the outgoing nucleon and instead
nucleon induced decays become dominant,
\begin{eqnarray}
\Lambda + p &\to& n + p ~\ : ~\ {\mbox{\rm proton induced decay}} \\
\Lambda + n &\to& n + n ~\ : ~\ {\mbox{\rm neutron induced decay}}
\end{eqnarray}
Such a process is regarded as a weak $\Delta S=1$ reaction of two 
baryons, which is analogous to the weak $NN$ interaction.
While the weak $NN$ interaction is masked by the strong interaction, 
the $\Lambda N \to NN$ transition provides us with a unique 
opportunity to study the mechanism of the baryon-baryon weak interaction.
Many theoretical and experimental studies have been done for 
the nonmesonic decays of light and heavy hypernuclei.
Yet mechanisms of the nonmesonic weak decay are not clear, 
especially as the predicted proton-induced decay rates are too
large and consequently the n/p ratio, 
the ratio of the neutron-induced decay rate, $\Gamma_{n}$,
to the proton-induced one, $\Gamma_{p}$, 
is too small compared with the observed value.

A conventional picture of $\Lambda N\to NN$ process 
is the one-pion exchange,
where $\Lambda N\pi$ vertex is induced by the weak 
interaction.\cite{OPE,ttb:ptp}
In $\Lambda N\to NN$, the relative momentum of the final state
nucleon is about 400 MeV/c, or (0.5 fm)$^{-1}$.
The nucleon-nucleon interaction at this momentum is dominated
by the short-range repulsion due to heavy meson exchanges and/or
to quark exchanges between the nucleons.\cite{QCM}
It is therefore expected that the short-distance interactions
will play significant roles in the two-body weak decay as well.
Exchanges of $K$, $\rho$, $\omega$, $K^*$ mesons and also correlated 
two pions have been studied,\cite{mcg:prc,dub:anp,par:prc,shm:npa}
and it is found that the kaon exchange is significant, 
while the other mesons contribute less.\cite{par:prc}

Several studies have been made on effects of quark 
substructure.\cite{chk:prc,mal:plb,iok:npa}
In our recent analyses,\cite{iok:npa,iom:npa,sasaki} 
we employ a $\Delta S= 1$ effective weak hamiltonian,
which consists of various four quark weak verteces
derived in the renormalization group approach to include QCD 
corrections to the $su\to ud$ transition mediated 
by the $W$ boson.{\cite{psw:npb}}
Such an effective weak hamiltonian has been applied to the decays of 
kaons and hyperons with considerable success.\cite{dgh:pre}  
Our aim is to explain the short range part of the weak baryonic 
interaction by using the same interaction so that we are able 
to confirm the consistency between the free hyperon decays and 
decays of hypernuclei.

We proposed to evaluate the effective hamiltonian in the six-quark 
wave functions of the two baryon systems and derived the 
``direct quark'' (DQ) weak transition potential for 
$\Lambda N \to NN$.\cite{iok:npa,iom:npa}
Our analysis shows that the direct quark contribution
largely improves the discrepancy between the meson-exchange theory
and experimental data for the n/p ratios.
It is also found that the $\DI=3/2$ component of the effective hamiltonian
gives a sizable contribution to $J=0$ transition amplitudes,\cite{mal:plb}
although we cannot determine the $\DI=3/2$ amplitudes 
unambiguously from the present experimental data.

\section{Direct quark mechanism}

In our recent papers,{\cite{iok:npa,iom:npa,sasaki}} we derived the
$\Lambda N \to NN$ transition potentials in the direct quark (DQ) 
mechanism. 
The transition potential is calculated by evaluating the effective 
hamiltonian in the nonrelativistic valence quark model.\cite{iok:npa}
The valence quark wave function of two baryon system is taken from 
the quark cluster model, which takes into account the 
antisymmetrization among the quarks.
Then the obtained transition potential has been applied to 
the weak decay of light hypernuclei.

In ref~\cite{iok:npa}, the DQ transition potential is calculated 
for all the two-body channels with the initial 
$\Lambda N (\ell=0)$ and the final $NN (\ell'=0,1)$ states.
The explicit forms of the transition potential are given in the 
momentum space representation. 
In ref.{\cite{sasaki}}, it is expressed in the coordinate space formulation, 
so that realistic nuclear wave functions with short range 
correlations can be easily handled.
The DQ transition potentials in the coordinate space contain
nonlocal terms as a result of the quark antisymmetrization.
It also has terms with a derivative operator acting on the initial 
relative coordinate.
Thus the general form of the transition potential is 
\begin{eqnarray}
 {V_{DQ}}^{\ell\ell'}_{ss'J}(r,r') &=& 
 \langle NN: \ell' s' J|V({\vec r'},
 {\vec r}) |\Lambda N : \ell s J\rangle \nonumber\\
  &=& V_{loc}(r)\, {\delta(r-r')\over r^{2}}
    + V_{der}(r)\, {\delta(r-r')\over r^{2}} \partial_{r}
    +V_{nonloc} (r,r')
\label{dqpot}
\end{eqnarray}
where $r$ ($r'$) denotes the relative radial coordinate of the 
initial (final) two-baryon state, and $\partial_r$ stands for the 
derivative of the initial $\Lambda N$ relative wave function.

We later combine the direct quark (DQ) and meson exchange (ME) potentials, 
giving the total transition potential as
\begin{equation}
V(r,r') = V^{ME}(r) {\delta(r-r')\over r^2} + V^{DQ}(r,r')
\end{equation}
The relative phase between ME and DQ is fixed so that the weak quark 
hamiltonian gives the correct amplitude of $\Lambda\to N\pi$ 
transition.{\cite{iom:npa}}  
Note that the relative phases among various meson exchange potentials 
are determined according to the $SU(6)_w$ symmetry.

\section{Meson exchange mechanism}
The one pion exchange (OPE) process is expected to give a main part 
of the $\Lambda N \to NN$ transition 
potential.  The weak $\pi N \Lambda$ vertex is parametrized as
\begin{equation}
{\cal{H}}_w^{\pi} =
         iG_Fm_{\pi}^2 \overline{\psi}_N ( A_\pi + B_\pi \gamma_5 ) 
         \vec{\tau} \cdot \vec{\phi}_\pi \psi_\Lambda.
       \label{eqn:phenoweak}
\end{equation}
where the coupling constants, $A_\pi$ and $B_\pi$,
are determined from the $\Lambda \to N\pi$ decay amplitudes.
As the momentum transfer in this transition is fairly large, 
the tensor part plays an important role.  As a result, the
$\Lambda p(^3S_1)$ to $pn(^3D_1)$ transition becomes dominant.
This tensor dominance causes the n/p ratio problem, namely,
the observed n/p ratio ($\simeq 1$) can not be explained
by the OPE transition.

While the OPE is significant
for the long-distance baryon-baryon interaction, 
the short range reaction mechanism is also important 
in $\Lambda N \to NN$
due to the large energy transfer involved.
Within the meson exchange model, a shorter range contributions 
may come from the exchange of the heavier mesons
\cite{mcg:prc,dub:anp,par:prc}
and multi-correlated meson.\cite{shm:npa}
In ref.\ \cite{par:prc}
the authors consider the exchange of all the octet pseudoscalar 
and vector mesons, 
$\pi$, $K$, $\eta$, $\rho$, $\omega$, and $K^\ast$.
While the $\Lambda N \pi$ weak coupling constant 
is determined phenomenologically, 
all the other weak couplings in ref.\ \cite{par:prc}
are only estimated theoretically by assuming
the $SU(6)_w$ symmetry.\cite{dub:anp}
For the strong vertices, 
the coupling constants are taken from the Nijmegen YN potential 
model D.

\begin{figure}[tb]
\centerline{ \epsfxsize=12cm \epsfbox{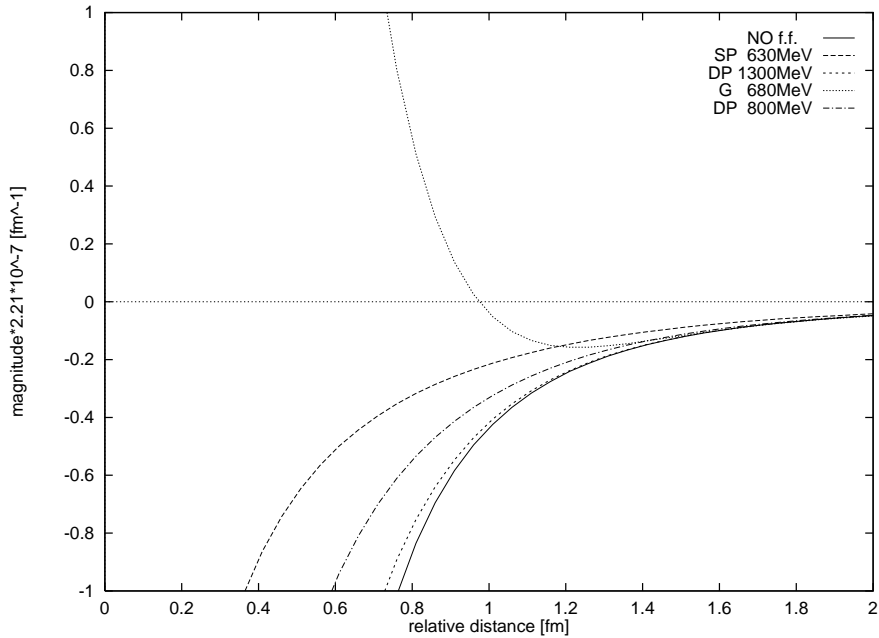} }
\caption{Potentials for the $^3S_1-^3D_1$, with relative momentum of 
$k_r=1.97$fm$^{-1}$}
\label{pot9}
\end{figure}

As the baryons and mesons have finite size and structure,
we consider a form factor at each meson-baryon vertex.
If we assume the same form factor $F({\bf q}^2)$ 
both for the strong and weak vertices,
the potential is given as
\begin{equation}
V ({\bf{r}}) 
    = \int \frac{d^3q}{(2\pi)^3} \frac{e^{i {\bf{q \cdot r}}}}
   {{\bf{q}}^2 + \mu^2 - q_0^2} {\cal{O}} ({\bf{q}}) F^2({\bf q}^2)
   \label{ffp}
\end{equation}
where ${\cal O}$ is a product of vertex operators and coupling constants.
A standard choice of $F^2({\bf q}^2)$ is 
the square of the monopole form factor,
\begin{equation}
F^2_{DP}({\bf{q}}^2) 
    = \left( \cfrac{\Lambda_{DP}^2 - \mu^2}{\Lambda_{DP}^2 
         + {\bf{q}}^2} \right)^2
\end{equation}
which we call ``double  pole'' (DP) form factor.
On the other hand, a simpler form is often used in literatures,
\begin{equation}
F^2_{SP}({\bf{q}}^2) 
    = \cfrac{\Lambda_{SP}^2 - \mu^2}{\Lambda_{SP}^2 + {\bf{q}}^2}
\end{equation}
which we call ``single pole'' (SP) form factor.
The cutoff parameter $\Lambda$ 
is chosen for individual mesons independently.
Another type of form factor considered here is the ``Gaussian'' (G),
\begin{equation}
F^2_G({\bf{q}}^2) 
    = \exp \left( -\frac{{\bf{q}}^2}{\Lambda^2} \right),
\end{equation}
which has the advantage of the consistency 
with the quark structure for the baryon, 
if the cutoff parameter is taken according to 
the size of quark distribution in the baryon.
While the SP and DP form factors give similar effects,
the G form factor behaves differently at short distances.
Fig.~\ref{pot9} shows the behaviors of the OPE potential  
in the $\Lambda N(^3S_1)$ - $NN(^3D_1)$ channel
at the relative momentum of $k_r = 1.97$fm$^{-1}$.
As can be seen clearly, the amplitude has a node for the G form factor, 
while the others do not behave like that.
This oscillation suppresses the tensor transition 
although the long distance behavior is similar to the ``hard'' 
DP form factor or that without the form factor.
We test these three form factors and compare the results.

We consider the effect of the finite energy transfer $q_0$ in eq.\ 
(\ref{ffp}).
Suppose that the initial $\Lambda-N$ pair is at rest, then
the energy transfer is given by
$
q_0 \simeq (m_\Lambda - m_N)/2 \simeq 88{\mbox{\rm MeV}}.
$
Taking this $q^0$ as an average, we
introduce the effective mass of the meson,
$
\tilde{\mu}_i = \sqrt{\mu_i^2 - (m_\Lambda - m_N)^2/4}
$
and make the following replacement,
\begin{equation}
\frac{1}{{\bf{q}}^2 + \mu^2 - q_0^2} \to \frac{1}{{\bf{q}}^2 + \tilde{\mu}^2}
\end{equation}
This effect leads to measurable changes.
For example, the pion mass is reduced by about $25\%$,
and the range of OPE becomes longer.

\section{Decay of $\Lambda$ in nuclear medium}

First we study the $\Lambda$ decay in nuclear matter 
in order to investigate our approach to $\Lambda N \to NN$
without ambiguities from the nuclear structure.
It is regarded as an approximation of decays of heavy hypernuclei.
Recent experimental data of the life-times of heavy hypernuclei indicate 
that the nonmesonic decay rate of $\Lambda$ in heavy hypernuclei 
is only $20\%$ larger than the free $\Lambda$ decay rate.\cite{pby:psg}
This saturation suggests 
the short-range nature of the $\Lambda N \to NN$ decay.

We assume that nuclear matter is symmetric ($N_n = N_p$) 
and the $\Lambda$ is at rest in it.
We employ the plane wave with the short range correlation
for the wave function of both the initial and final states. 
The correlation function, $f(r)$, proposed in ref.~\cite{par:prc},
is multiplied to the relative wave functions.
For the initial $\Lambda N$ states, we employ
\begin{equation}
f_i(r) = \left( 1-e^{-\frac{r^2}{a^2}} \right)^n + br^2 e^{-\frac{r^2}{c^2}}
\end{equation}
where $a=0.5$, $b=0.25$, $c=1.28$, $n=2$, and for the final $NN$ 
state,
\begin{equation}
f_f(r) = 1-j_0( q_c r ) 
\end{equation}
where $q_c=3.93$ fm$^{-1}$.
The initial state correlation is obtained 
from a macroscopic finite nucleus G-matrix calculation, 
and the final state correlation 
gives a good description of nucleon pairs in $^4$He.

In the present calculation, we only consider the relative $L = 0$ 
for the initial $\Lambda - N$ system.  It has been pointed out by 
explicit calculation that contributions of the higher partial waves 
are of 5\% order.
The decay rate is given by
\begin{eqnarray}
\Gamma_{N M} &=&
       \frac{m_N}{(2\pi)^5 \mu^3} \int^{\mu k_F}_0 k^2 k_0 dk \\
       &\times&\frac{1}{4} \sum_{p,n}
       \left( |a|^2 + |b|^2 + 3|c|^2 + 3|d|^2 + 3|e|^2 + 3|f|^2 \right)
\end{eqnarray}
where
$k_0^2 \equiv m_N ( m_\Lambda - m_N ) + k^2/2\mu $
and 
$\mu \equiv m_\Lambda/(m_N + m_\Lambda)$.
The $a$, $b$, $\cdots$, $f$ are the matrix elements of 
the transition potential in each channel designated by its final orbital 
angular momentum, the initial and final spins.

\subsection{Form factors}
As the $\Lambda N \to NN$ scattering involves a large momentum transfer, 
$\sim 400MeV/c$, 
the amplitudes are rather sensitive to the form factor.
Here we study choices of the form factor for OPE.
Table~\ref{ffc} shows the calculated decay rates of $\Lambda$ in OPE.
The SP form has often been used in literatures.\cite{mcg:prc,ttb:ptp}
Ref.{\cite{mcg:prc}} employed the SP form factor 
with $\Lambda_{SP}^2 = 20m_\pi^2$, or $\Lambda_{SP} \sim 630$MeV.
This form factor is based on the dispersion relation analysis 
with the semi-pole approximation to the $\pi NN$ form factor.
It is a very soft form factor, 
which cuts off the short-range part rather drastically
and the resulting decay rate becomes small.
The tensor transition, $\Lambda p(^3S_1) \to pn(^3D_1)$, 
seems most affected by such form factor. 
We find that the tensor transition amplitudes 
are reduced by a factor two or more by the form factor.
Therefore total decay rate is much reduced.

\begin{table}[ht]
\caption{Decay rates of $\Lambda$ in nuclear matter 
         (in units of $\Gamma_\Lambda = (263 \times 10^{-12}sec)^{-1}$)
         for various choices of the form factors
         in OPE.}
\begin{center}
\begin{tabular}{||c||ll|c|c|c|c|c||} 
\hline
 & & & $total$ & $\Gamma_p$ & $\Gamma_n$ & ${\Gamma_n}/{\Gamma_p}$ & $PV/PC$ \\
\hline
$\pi$ & no-f.f.&                    & 2.819 & 2.571 & 0.248 & 0.097 & 0.358 \\
      & SP & $\Lambda_\pi = 630$MeV & 1.103 & 0.989 & 0.114 & 0.116 & 0.408 \\
      &    & $\Lambda_\pi = 920$MeV & 2.332 & 2.116 & 0.216 & 0.102 & 0.376 \\
      & DP & $\Lambda_\pi =1300$MeV & 2.575 & 2.354 & 0.221 & 0.094 & 0.337 \\
      &    & $\Lambda_\pi = 800$MeV & 1.850 & 1.702 & 0.148 & 0.087 & 0.282 \\
      & G  & $\Lambda_\pi = 680$MeV & 1.514 & 1.129 & 0.386 & 0.342 & 0.580 \\
 \hline
\end{tabular}
\end{center}
\label{ffc}
\end{table}

The DP form factor is popular in meson exchange potential models 
of the nuclear force.
The Bonn potential, for instance, employs the DP 
with $\Lambda_{DP}(\pi) = 1300$MeV.
This form factor is extremely hard so that 
the short range part of the meson exchange potential becomes relevant.
Lee and Matsuyama carried out an analysis of the $NN \to NN\pi$ processes 
and pointed out \cite{lee:prc} that a soft form factor, 
such as $\Lambda<750$MeV, is preferable.
Recent analysis in the QCD sum rule 
also suggests a soft $\pi NN$ form factor of cutoff 
$\Lambda \simeq 800$[MeV]~\cite{mei:prc}.
From the quark substructure point of view, 
it is also natural that the cutoff $\Lambda$ is 
of order $\sim 1/R_h \sim 400 - 500$MeV.
In Table~\ref{ffc}, we find that the soft form factor reduces 
the decay rates by about $40\%$, 
while the hard one changes only by $\le 10\%$.

The G form factor is related to the Gaussian quark wave function of the baryon.
Corresponding to the size parameter $b \simeq 0.5$fm, 
we employ $\Lambda_G(\pi) \simeq 680$MeV 
because the relation between the size and cutoff parameters 
is $\Lambda =\sqrt{3}/b$.

The above sensitivity to choice of the form factor is quite annoying 
because it is not easy to judge which of the form factors is the right one.
It should also be noted that the weak vertex form factor 
can be different from the strong one, 
although the pole dominance picture of the parity-conserving weak vertex 
leads to the identical form factor to the strong one.
In the meson exchange potential models of the nuclear force, 
there seems a tendency to choose a hard form factor
because the heavy meson exchanges 
are often important for spin-dependent forces.
On the other hand, the quark model approach to the short range nuclear force 
gives significant spin dependencies comparable to the heavy meson exchanges 
and therefore the meson exchanges can be cut off rather sharply 
with a soft form factor.
In the present approach to the weak $\Lambda N \to NN$ interaction, 
whose short range part is described by the direct quark mechanism, 
we thus may follow the quark model approach to the nuclear force 
and take the ``soft'' form factor for OPE potential as a standard.

\subsection{Decay rates in nuclear matter}

\begin{table}[tbh]
\caption{Nonmesonic decay rates of $\Lambda$ in nuclear matter 
          (in units of $\Gamma_\Lambda$). 
          The ``all'' includes 
          $\pi$, $K$, $\eta$, $\rho$, $\omega$, and $K^\ast$ meson exchanges.
	  The ``DP hard (soft)'' assumes $\Lambda_{\pi}=1300 (800)$ MeV. 
}
 \begin{center}
 \begin{tabular}{||cl||c|c|c|c|c||} 
 \hline
  & & $total$ & $\Gamma_p$ & $\Gamma_n$ & ${\Gamma_n}/{\Gamma_p}$ & $PV/PC$ \\
 \hline
 $\pi$ &DP hard     & 2.575 & 2.354 & 0.221 & 0.094 & 0.337 \\
       &DP soft     & 1.850 & 1.702 & 0.148 & 0.087 & 0.282 \\
       &no-f.f.     & 2.819 & 2.571 & 0.248 & 0.097 & 0.358 \\
 \hline
 $\pi$+K &DP hard     & 1.099 & 1.075 & 0.024 & 0.022 & 0.631 \\
         &DP soft     & 0.695 & 0.674 & 0.021 & 0.031 & 0.632 \\
         &no-f.f.     & 1.143 & 1.111 & 0.032 & 0.028 & 0.745 \\
 \hline
 all &DP hard     & 1.672 & 1.571 & 0.101 & 0.064 & 1.468 \\
     &DP soft     & 1.270 & 1.152 & 0.117 & 0.101 & 1.537 \\
     &no-f.f.     & 1.744 & 1.704 & 0.040 & 0.024 & 2.952 \\
 \hline
 DQ &                       & 0.418 & 0.202 & 0.216 & 1.071 & 6.759 \\
 \hline
 DQ+$\pi$ &DP hard & 3.609 & 2.950 & 0.658 & 0.223 & 0.856 \\
          &DP soft & 2.726 & 2.202 & 0.523 & 0.238 & 0.896 \\
 \hline
 DQ+$\pi$+K &DP hard & 1.766 & 1.495 & 0.271 & 0.181 & 1.602 \\
            &DP soft & 1.204 & 0.998 & 0.206 & 0.207 & 1.954 \\
 \hline
 \hline
\end{tabular}
\end{center}
\label{tot}
\end{table}

The calculated decay rates of $\Lambda$ in nuclear matter 
for several models are listed in Table~\ref{tot}.
We compare the results of the soft DP form factor 
and those of the hard DP form factor.
One sees that the even the soft-cutoff OPE predicts 
a total decay rate much larger than the observed one, about 
$1.2\times\Gamma_{\Lambda}$, 
for the heavy hypernuclei.
This is due to the large
tensor transition $d_p$, $\Lambda p(^3S_1) \to pn(^3D_1)$, 
which is allowed only in the proton-induced decay.
This tensor dominance property leads to 
the small $\Gamma_n/\Gamma_p$ and $PV/PC$ ratio.

The kaon exchange contribution 
reduces $\Gamma_p$ by more than factor two.
This mainly comes from the cancelation in the channel $d_p$.
At same time, the kaon exchange contribution reduces 
$\Gamma_n$.
Therefore, the $n/p$ ratio remains small.
The decay is dominated by the $J=1$ channels.

When we include $\eta$, $\rho$, $\omega$, and $K^\ast$ (``all'') mesons,
both the proton and neutron induced decays increase 
but the $n/p$ ratio is still small ($\simeq 0.1$).
It is interesting to see that the PV/PC ratio becomes large,
when we include heavier mesons.

One sees in Table~\ref{tot} that 
the magnitude of DQ itself is small compared to that of $\pi$ or $\pi+K$.
However, the DQ has a large $\Gamma_n$
and a large $n/p$ ratio.
It is also shown that the DQ mechanism is dominated by 
the parity violating channels and thus produces a large $PV/PC$ ratio.
Such charasteristic behaviors of DQ 
will distinguish DQ from the other mechanisms.

In $DQ+$meson exchanges, the pion contribution is again senseitive to 
the choice of the form factor.
One sees that the softer form factor is more appropriate 
to reproduce experimental values of $\Gamma_p$.
As the DQ has little contribution to the tensor channel, 
$\Gamma_p$ is still too large in ``$DQ+\pi$''.
Again, the $K$ exchange reduces the tensor amplitude
and thus $\Gamma_p$ is suppressed by a factor 2.
However $\Gamma_n$ is reduced, at the same time, 
which results in the $n/p$ ratio$\simeq 0.2$.
This value is still smaller than the observed values
for the medium and heavy hypernuclei.

Contribution beyond the $\pi$ and $K$ mesons
does not improve the situation.
It is also questionable whether the DQ mechanism 
and the vector meson exchanges are independent and can be superposed.
The double counting problem for the vector mesons 
and DQ contribution in nuclear force is pointed out in ref \cite{yaz:ppn}.
We thus take the ``$DQ+\pi+K$'' with the soft $\pi$ form factor
as our present best model for the nonmesonic $\Lambda$ decay.

\subsection{Decays of light hypernuclei}
The same $\Lambda N \to NN$ transition potential 
has been applied to the study of the nonmesonic weak decays 
of light s-shell hypernuclei,
${^5_\Lambda He}$, ${^4_\Lambda He}$, and ${^4_\Lambda 
H}$.{\cite{sasaki}}
The results are summarized in Table~\ref{lig}.

%
%
%
%

\begin{table}[thb]
\begin{center}
\caption{Nonmesonic decay rates (in units of $\Gamma_\Lambda$)
         of light hypernuclei. 
         The DP (soft) form factor is used for OPE.}
\begin{tabular}{||cl|c|c|c|c||} 
\hline
 & & $total$ & $\Gamma_p$ & $\Gamma_n$ & ${\Gamma_n}/{\Gamma_p}$ \\
\hline
${^5_\Lambda He}$ &  $\pi$   & 0.740 & 0.654 & 0.087 & 0.133 \\
                  & $\pi+K$  & 0.350 & 0.331 & 0.028 & 0.055 \\
                  &$\pi+K+DQ$& 0.521 & 0.435 & 0.085 & 0.195 \\
\cline{3-6}
                  & exp.~\cite{jjs:prc} & 0.41$\pm$0.14 & 0.21$\pm$0.07 &
                               0.20$\pm$0.11 & 0.93$\pm$0.55 \\
                  & exp.~\cite{hno:psg} & 0.50$\pm$0.07 & 0.17$\pm$0.04 &
                               0.33$\pm$0.04 & 1.97$\pm$0.67 \\
 \hline
\hline
${^4_\Lambda He}$ &  $\pi$   & 0.542 & 0.498 & 0.044 & 0.089 \\
                  & $\pi+K$  & 0.252 & 0.233 & 0.019 & 0.082 \\
                  &$\pi+K+DQ$& 0.309 & 0.302 & 0.007 & 0.024 \\
\cline{3-6}
                  & exp.~\cite{hno:psg} & 0.19$\pm$0.04 & 0.15$\pm$0.02 
		  & 0.04$\pm$0.02 & 0.27$\pm$0.14 \\
 \hline
\hline
${^4_\Lambda H}$  &  $\pi$   & 0.080 & 0.022 & 0.056 & 2.596 \\
                  & $\pi+K$  & 0.020 & 0.010 & 0.010 & 1.099 \\
                  &$\pi+K+DQ$& 0.120 & 0.060 & 0.059 & 0.983 \\
\cline{3-6}
                  & exp.~\cite{hno:psg} & 0.15$\pm$0.13 & ----- 
		  & ----- & ----- \\
 \hline
\hline
\end{tabular}
\end{center}
\label{lig}
\end{table}

We find that the ``$\pi$ + $K$ + DQ'' picture 
gives us a fair agreement with experiment for the total decay rate,
while the n/p ratios are still too small.
As seen in the nuclear matter calculation, 
the main contribution comes from the OPE mechanism, 
which produces the large proton-induced rate.
Comparing with the experimental data 
we find that the proton-induced rate is overestimated in all the pictures, 
while the neutron-induced rate is underestimated.

\section{$\pi^+$ decays and $\DI=3/2$ amplitudes}
\def\Lam{\Lambda}
\def\Sig{\Sigma}
\def\DI{\Delta I}
\def\pip{\pi^+}
\def\pim{\pi^-}
\def\piz{\pi^0}
\def\vs{{\it v.s.}}
\def\ie{{\it i.e.}}

The pionic decay is another interesting decay mode of light hypernuclei.
While the free $\Lam$ 
decays into $p\pim$ or $n\piz$, the $\pip$ decay requires an assistance
of a proton, \ie, $\Lam+p\to n+n+\pip$.
Some old experimental data suggest that the ratio of $\pip$ and 
$\pim$ emission from $^4_{\Lambda}He$ is about 5\%.{\cite{pip-data}}
The most natural explanation of this process is $\Lam\to n\pi^0$ decay 
followed by
$\pi^0 p \to \pip n$ charge exchange reaction.
It was evaluated for realistic hypernuclear wave functions and
found to explain only 1.2\% for the $\pip/\pim$ ratio.{\cite{DHCG}}
Another possibility is to consider $\Sig^+ \to \pip n$ decay
after the conversion $\Lam p \to \Sig^+ n$ by the strong interaction
such as pion or kaon exchanges.
It was found, however, that the free $\Sig^+$ decay which is dominated
by $P$-wave amplitude, gives at most 0.2\% for the $\pip/\pim$ ratio.
Recently, it was proposed that a two-body process $\Sig^+ N\to n N\pip$
must be important in the $^4_{\Lambda}He$ decay.{\cite{GT}}
But its microscopic mechanism is not specified.

\let\di=\DI

In order to solve this problem, we have applied 
the soft-pion technique to the $\pip$ decay of light
hypernuclei.{\cite{piplus}}
The soft-pion theorem to the process $\Lam p \to nn \pip(q\to 0)$
reads
\begin{equation}
  \lim_{q\to 0} \langle nn\pip(q)|H_W|\Lam p\rangle 
= -{i\over \sqrt{2} f_{\pi}} \langle nn|[Q_5^-, H_W]|\Lam p \rangle
\label{soft-pion}
\end{equation}
Again, because of
\begin{equation}
  [Q_5^-, H_W] = -[ I_-, H_W]
\end{equation}
it discriminates the isospin properties of $H_W$.
Similarly to the case of $\Sig^{+}$ decay, we see
that the $\di=1/2$ part vanishes as
\begin{eqnarray}
  [I_-,H_W(\di=1/2, \di_z=-1/2)] &=& 0 \\{}
  [I_-,H_W(\di=3/2, \di_z=-1/2)] &=& \sqrt{3} H_W(\di=3/2, 
  \di_3=-3/2) 
\end{eqnarray}
We then obtain
\begin{equation}
  \lim_{q\to 0} \langle nn\pip(q)|H_W|\Lam p\rangle 
= {i\sqrt{3} \over \sqrt{2} f_{\pi}} 
   \langle nn|H_W(\di=3/2, \di_3=-3/2)|\Lam p \rangle
   \label{eq:pipamp}
\end{equation}
Thus we conclude that the soft $\pip$ emission in the 
$\Lam$ decay in hypernuclei is caused only by the $\di=3/2$ 
component of the strangeness changing weak hamiltonian.
In other words, the $\pip$ emission from hypernuclei probes the 
$\di=3/2$ transition of $\Lam N \to NN$.

\section{Future prospects}

The $\Lambda$ decays iin nuclear medium is a unique reaction as a 
weak interaction which involves two baryons.  Its study may reveal 
roles of quark substructures in weak processes and the mechanism of 
the $\DI=1/2$ enhancement in nonleptonic weak decays of strange 
hadrons.  There are many unknowns, such as the weak meson-baryon 
coupling constants, the validity of $SU(6)_{w}$ symmetry, the 
magnitude of the parity violation, and so on.
Both experimental and theoretical efforts may produce a lot of useful 
and interesting information in understanding hadronic weak interactions.

Here we considered the new direct quark (DQ) mechanism for the 
$\Lambda N \to NN$ process, which induces decays of hypernuclei.
Combining the meson exchange processes for a long range part, 
we find that the ``$\pi + K + DQ$'' mechanism
with soft DP form factor for OPE 
explains the total decay rates of both light and heavy hypernuclei 
rather well.  
There, however, remains a difficult problem.
That is, the proton-induced decay rate is overestimated
and thus we predict a small n/p ratio.
The ratio is imporoved by the DQ contribution.
It is yet too small ($\simeq 0.2$) for ``$\pi + K + DQ$''.
The experimental values are not completely fixed,
but they suggest $\simeq 1$ for heavy hypernuclei.
The situation is similar for $^5_\Lambda$He.
It is thus urgent and important to find out 
what causes this discripancy.

\def \vol(#1,#2,#3){{{\bf {#1}} (19{#2}) {#3}}}
\def \NP(#1,#2,#3){Nucl.\ Phys.\          \vol(#1,#2,#3)}
\def \PL(#1,#2,#3){Phys.\ Lett.\          \vol(#1,#2,#3)}
\def \PRL(#1,#2,#3){Phys.\ Rev.\ Lett.\   \vol(#1,#2,#3)}
\def \PRp(#1,#2,#3){Phys.\ Rep.\          \vol(#1,#2,#3)}
\def \PR(#1,#2,#3){Phys.\ Rev.\           \vol(#1,#2,#3)}
\def \PTP(#1,#2,#3){Prog.\ Theor.\ Phys.\ \vol(#1,#2,#3)}
\def \ibid(#1,#2,#3){{\it ibid.}\         \vol(#1,#2,#3)}
\def\MO{M.~Oka} \def\etal{{\it et al.}}

\end{document}